\documentclass[pre,aps,eqsecnum,showpacs,draft]{revtex4} 
\input{epsf}
\begin{document}

\title{NUMERICAL STUDY OF THE SHAPE AND INTEGRAL PARAMETERS OF A DENDRITE}

\author{R. Gonz\'alez-Cinca and L. Ram\'{\i}rez-Piscina}

\affiliation{ Departament de F\'{\i}sica Aplicada, Universitat Polit\`ecnica de
Catalunya, Av. del Canal Ol\'{\i}mpic s/n, 08860 Castelldefels 
(Barcelona), SPAIN\\
Tel.: (+34) 934137082, fax: (+34) 934137007,  e-mail: ricard@fa.upc.es }

\date{\today}

\begin{abstract}

We present a numerical study of sidebranching of a solidifying dendrite by means
of a phase--field model. Special attention is paid to the regions far from the tip
of the dendrite, where linear theories are no longer valid. Two regions have been
distinguished outside the linear region: a first one in which sidebranching is in
a competition process and a second one further down where branches behave as
independent of each other. The shape of the dendrite and integral parameters
characterizing the whole dendrite (contour length and area of the dendrite) have
been computed and related to the characteristic tip radius for both surface
tension and kinetic dominated dendrites. Conclusions about the different
behaviors observed and comparison with available experiments and theoretical
predictions are presented.

\end{abstract}


\maketitle


\section{Introduction}

The generation of dendritic patterns 
arises in different nonequilibrium situations.
\cite{langer87,pelce,benjacob,brener91,godreche,hurle,santafe,langer83}.  
The case 
of dendrites appearing during the solidification of a melt has long provided an
archetypical example of pattern forming system, in which the underlying physics
is well known. Nonetheless, from the theoretical point of view it has posed a
number of nontrivial questions on the selection of the final growth mode, which
in a large amount has driven the research on the effects of nonlinearities,
anisotropies and fluctuations in interfacial pattern formation. These questions
also have an applied interest, since
solidification is one of the most common methods to produce materials.
It is well known that the details of the dendritic pattern (and in particular its
associated scales) appearing during growth, determine the microstructure of the
grown solid, 
which in turn is responsible to a large degree of its final (mechanical and
electrical) properties \cite{acta00}.

In this context an increasing attention is focussed on the shape of the growing
dendrite and on sidebranching
\cite{pieters86,martin,barber,langer87b,pieters88,vansaarloos,brener95,karma99,
pavlik99,pavlik00,rgc01,dougherty87,couder,qian,bouissou,dougherty92,
hurlimann,williams,bisang95,bisang96,li98,li99,lacombe,rgccou,courgc1,courgc2,
rgc02,rgc02b}, 
which corresponds to the apparition and growth of secondary branches at both sides
of the dendrite.
Sidebranching activity shows different behaviors
depending on the distance to the tip of the dendrite
\cite{brener95,courgc1,courgc2}. In the zone closer to the tip sidebranches are
born as a convective instability of the dendrite and grow linearly. Further down
from the tip, sidebranches are usually much more developed and a competition
process between branches takes place mediated by the interaction between their
diffused fields. Much further from the tip, the competition has finished and
winner branches grow as free dendrites while the growth of looser branches is
inhibited.

Many theoretical
\cite{barber,langer87b,pieters88,vansaarloos,brener95,karma99,pavlik99,
pavlik00,rgc01} and experimental
\cite{dougherty87,couder,qian,bouissou,dougherty92,hurlimann,williams,
bisang95,bisang96,li98,li99,lacombe,rgccou,courgc1,courgc2} 
studies of the region close to the tip 
have been carried out in the recent years. A common point in the study of the
linear region has been the characterization of sidebranching by means of its
amplitude and wavelength. 
The growth of sidebranches in the regions further down from the tip presents a
behavior very different from that of the linear regime. One finds first a region
where branches compete interacting through the expelled heat. This gives rise to
an irregular growth of sidebranches which is difficult to characterize by their
amplitude and wavelength, since these quantities are not longer well defined
outside the linear region. 
Experiments carried out with different substances
\cite{dougherty92,hurlimann,li98} have shown that sidebranching in this region is
self-similar and that geometrical parameters can be scaled by the tip radius $R$.
This nonlinear region and its associated self-similar growth have their limits at
distances to the tip of the order of the diffusion length, {\it i.e.}  $z/R <<
1/Pe$ \cite{brener95}, where $Pe$ is the P\`eclet number. Further down,
sidebranches behave like dendrites themselves.

The nonlinear region has been intensively studied in experiments with xenon
dendrites by H\"{u}rlimann {\it et al.} \cite{hurlimann} and with succinonitrile
dendrites by Li and Beckermann \cite{li98,li99}. In particular, the shape of the
sidebranching envelope was studied in Ref. \cite{li98} by
measuring the distance $X$ from the axis of the dendrite to the tip of active
sidebranches (defined as those branches longer than all the other branches closer
to the tip) versus the distance $Z$ to the tip along the axis of the dendrite (see
Fig.~\ref{fig1}). Values of $X$ and $Z$ were computed from the image of the
dendrite projected on a plane, and far from the tip the relation
$X/R=0.668(Z/R)^{0.859}$ was obtained.

It was proposed in Refs. \cite{hurlimann,li98} an alternative set of integral
parameters in order to describe the complex shape of a dendrite as a whole and the
nonlinearities of dendritic solidification. Parameters characterizing independent
parts of the dendrite ({\it e.g.} amplitude and wavelength of the sidebranching)
do not take into account the interaction of the sidebranches through the diffusion
field. Nonlinear effects such as {\it e.g.} coarsening make unclear which
sidebranches should be included in the measurement of the wavelength and which
others should not. Instead, the contour length $U$, the projection area $F$ and
the volume of a dendrite appear to be more appropriate.

It was found in the earlier experimental work of Refs. \cite{hurlimann,li98} that
the projection area varied linearly with the contour length and the
corresponding slope $M$ accomplished $M/R=\text{\it constant}$. 
Similar results were found in early simulations \cite{rgc02}, but 
this was shown to be an effect of reflecting boundary conditions strongly
affecting the dendrite \cite{rgc02b}. This suggests that the experimental
observations could have been affected by the diffusion field of other close
dendrites or other growing morphologies. In more recent experimental \cite{li99}
and numerical \cite{rgc02b} works it was obtained that $F/(UR)$ was not a constant
in the nonlinear regime.

The projection area showed two different behaviors
($F/R^{2}=0.847(Z/R)^{1.598}$ for $Z/R<30$ and
$F/R^{2}=0.578(Z/R)^{1.72}$ for $Z/R>30$) in three-dimensional succinonitrile
dendrites \cite{li99}. However, in two-dimensional ammonium bromide dendrites
\cite{couder}, the area $F$ was found to vary over three orders of magnitude as
$Z^{1.5}$, as it would have happened if dendrites had had a smooth parabolic
shape. As regards the variation of the contour length with the distance to the
tip, only data corresponding to three-dimensional dendrites are available, where
two behaviors are distinguished
($U/R=0.887(Z/R)^{1.116}$ for $Z/R<20$ and $U/R=0.378(Z/R)^{1.50}$ for $Z/R>40$)
\cite{li99}.

An additional question is whether strong undercoolings can produce qualitative
changes in sidebranching characteristics. It is well known that increasing
undercooling the growth can switch from a regime dominated by surface tension to
a regime dominated by kinetic effects. This was already predicted
theoretically in Ref. \cite{brener91}. When anisotropies of both effects favor
different directions, changes in the growth directions of both dendrite and
branches occur by changing undercooling. Even if these anisotropies are in the
same directions, the behavior of the tip radius and velocity can present abrupt
changes. A numerical 
evidence of such changes 
can be found in Ref. \cite{rgclip}. 

In this paper we present a study of sidebranching by means of a phase--field model
for moving solid--liquid interfaces
\cite{kobayashi,wheeler,karma96,katona,rgcth,karma01,ricard-cuerno}.
We consider sidebranching generated by selective amplification of fluctuations
near the tip of a free growing dendrite. In particular, we focus on the
nonlinear zone, including both the region where competition occurs and further
down where sidebranches behave as free growing dendrites. Characterization is
performed working out the shape of the dendrite by means of its envelope, and
calculating the integral parameters. We have varied  undercooling in a large
range, in particular reaching relatively high values of the undercooling.  This
has permitted on the one hand, due to reduction in diffusion length, to access the
region of free growing
sidebranches far from the tip, and on the other hand to reach the kinetic regime
of growth.

This paper is organized as follows:
In Sec.~\ref{models} we present the classical sharp--interface model that
characterizes a solidification system, the phase--field model and the numerical
procedure used in this work. In Sec.~\ref{results} we present the results  of
simulations. We particularize in the effect of varying undercooling, and in the
differences between different zones of the dendrite. Detailed
characterization of the whole dendrite is performed by its shape and by
computing the integral parameters. Finally, concluding remarks are presented in
Sec.~\ref{conc}.

\section{Model and numerical procedure}\label{models}

The free solidification of a pure substance can be described by the
sharp--interface model \cite{langer87}, which relies on the heat diffusion
equation together with two boundary conditions at the interface, namely heat
conservation and Gibbs--Thomson (local equilibrium) equation:
\begin{equation} \label{diff}
\frac{\partial T}{\partial t} = D \nabla^2 T,
\end{equation}
\begin{equation} \label{cons}
L\upsilon_{n} = D c_{p} [(\nabla _n T)_{S}- (\nabla _n T)_{L}], \end{equation} %
\begin{equation} \label{gibbs}
T_{\text{interface}} = T_{M}-{T_{M} \over L} [\sigma(\theta)+\sigma''(\theta)]
\kappa -{\upsilon_{n} \beta_k(\theta)}.
\end{equation}
In these equations $T$ is the temperature ($T_{M}$ being the melting one), $D$ is
the
diffusion coefficient ($D=k/c_{p}$, being $k$ the heat conductivity and $c_{p}$
the specific heat per unit volume), $L$ is the latent heat per unit volume,
$\upsilon_{n}$ is the normal velocity of the interface, $\nabla _n$ is the normal
derivative at the interface (S and L referring to solid and liquid
respectively), $\sigma(\theta)$ is the anisotropic surface tension (where $\theta$
is the angle between the normal to the interface and some
crystallographic axis) and $\kappa$ is the local curvature of the interface.
$\beta_k(\theta)$ is an anisotropic kinetic term, introduced into the
Gibbs--Thomson Eq.~\ref{gibbs} to account for a linear nonequilibrium correction.

The results of simulations presented below have been obtained by means of a
phase--field model. These kind of models have received increased attention during
the last years \cite{ricard-cuerno}. One of their main features is the
introduction of an additional non--conserved scalar order parameter or
phase--field $\phi$, whose time evolution equation is coupled with the heat
diffusion equation through a source term in order to take into account the
boundary conditions at the interface. The phase--field takes constant values in
each of the bulk phases (in our case, $\phi=0$ in the solid and $\phi=1$ in the
liquid) changing continuously between them over a transition layer, the
interfacial thickness $\epsilon$. 
The equations of the model are then constructed in such a way that they converge
to the sharp--interface dynamics of Eqs.~(\ref{diff},\ref{cons},\ref{gibbs}) in
the limit of vanishing $\epsilon$.
Hence this parameter controls the convergence to the sharp--interface limit.

The corresponding equations for the time evolution of the phase--field and the
dimensionless temperature can be written in the following form \cite{wheeler}: %
\begin{eqnarray}
\label{phasefield} {\epsilon^2 \tau(\theta)} \frac{\partial \phi}{\partial t} =
\phi(1-\phi) \left(\phi-{1 \over 2}+30 \epsilon \beta \Delta u
\phi(1-\phi)\right) \nonumber \\
-\epsilon^2 \frac{\partial}{\partial x} \left[ \eta(\theta) \eta^{'}(\theta)
\frac{\partial \phi}{\partial y} \right] +\epsilon^2 \frac{\partial}{\partial y}
\left[ \eta(\theta)\eta^{'} (\theta)\frac{\partial \phi} {\partial x} \right]
+\epsilon^{2} \nabla \left[ \eta^2(\theta) \nabla \phi \right]
\end{eqnarray}
\begin{equation}
\label{tempdiff} {\frac{\partial u}{\partial t}}+{1 \over \Delta} \left( 30
\phi^{2} - 60 \phi^{3} + 30 \phi^{4} \right) \frac{\partial \phi}{\partial t} =
\nabla^{2} u + \psi(x,y,t)
\end{equation}
where $u({\bf r},t) $ is the diffusion field and $\Delta = c_p \Delta T / L$ is
the dimensionless undercooling. Lengths are scaled by some arbitrary reference
length $\omega$, while times are scaled by $\omega^{2}/D$. In these equations
$\theta$ is the angle between the $x$-axis and the gradient of the phase--field.
$\eta(\theta)=\sigma(\theta)/\sigma(0)$ is the anisotropy of the surface tension.
$\tau(\theta)$ is given by $\frac{c_p D}{L d_0} \eta(\theta) \beta_k (\theta)$,
so the anisotropy of the kinetic term is given by $\tau(\theta)/ \eta(\theta)$.
$\beta$ is equal to $\frac{\sqrt{2} \omega}{12 d_o}$ and $d_o=c_p T_M \sigma(0)
/ L^2$ is the capillary length.

A source of fluctuations is introduced through the additive term $\psi$ in the
heat equation. It was demonstrated \cite{rgc01} that sidebranching induced by this
kind of noise qualitatively reproduce the characteristics of the
(thermodynamical) internal noise, which makes it appropriate for the study of
sidebranching. In our two-dimensional simulations the noise term is evaluated at
each cell $(i,j)$ of lateral size $\Delta x$ as $I
r_{ij}$, where $I$ denotes the amplitude of the noise, and $r_{ij}$ is an
uncorrelated uniform random number in the interval $[-0.5,0.5]$. The
phase--field model equations have been solved on rectangular lattices using
first-order finite differences on a uniform grid with mesh spacing $\Delta x$. An
explicit time-differencing scheme has been used to solve the equation for $\phi$,
whereas for the $u$ equation the alternating-direction implicit method was chosen
\cite{press:92}. 
The kinetic term has been taken as isotropic, which leads to
$\tau(\theta)=m\eta(\theta)$ with constant $m$. A four-fold surface tension
anisotropy $\eta(\theta) =1+\gamma \mathrm{cos}(4\theta)$ has been considered.

The growth morphologies have been obtained by setting a small vertical seed
($\phi=0$, $u=0$) in the center of the bottom side of the system and imposing
$\phi=1$ and $u=-1$ on the rest of the system. Symmetric boundary conditions for
$\phi$ and $u$ have been used on the four sides of the system. 
Special care has been taken to employ large enough system sizes to avoid any
influence of boundary conditions on the results presented along this paper.

We have used a set of phase--field model parameters that gives rise to a growing
needle without sidebranching when no noise ($I=0$) is added to the simulations.
This assures us that the sidebranching observed when $I \neq 0$ is not due to
numerical noise. The fixed parameters for all the simulations have been
$\beta=320$, $\gamma=0.045$, $m=16$ and $\epsilon=3.75 \times 10^{-3}$. The value
of $\Delta$ has been varied in the range $0.44-0.65$. The noise amplitude was kept
constant ($I=16$) in all the simulations and the time and spatial discretizations
used were $\Delta t=1.25 \times 10^{-4}$ and $\Delta x=0.0125$.

\begin{figure}
\begin{center}
\parbox{8cm}{
\epsfxsize=8cm \epsfbox{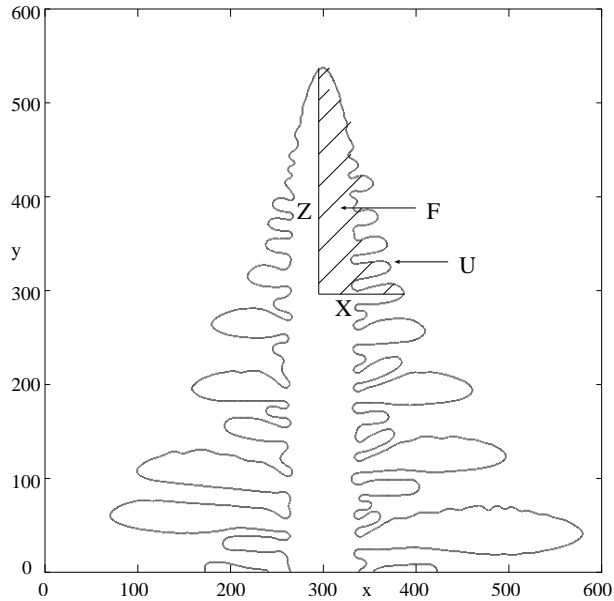}} \caption[]{\label{fig1} Example of dendrite
obtained at $\Delta=0.575$. Distances $Z$ and $X$ of active sidebranches used to
characterize the shape of the dendrite, and the integral parameters contour length
($U$) and surface area ($F$), are indicated.}
\end{center}
\end{figure}

Under these conditions, the obtained morphologies have been dendrites with three
main arms growing from the seed, one in the vertical ($y$) direction and two in
the horizontal ($x$) one. We have focussed on the sidebranches which grew
perpendicular to the vertical arm. Thus, in order to get rid of the influence of
the diffusion field of the horizontal arms on these sidebranches, we have been
forced to run long simulations and only observe an area at a fixed distance to the
tip. 
In order to avoid working with unnecessarily large systems, we have performed
periodic shifts of the complete system, practice that has been checked it did not
affect the results of the simulation.
If Fig.~\ref{fig1} we show an example of a typical grown dendrite.

\section{Results and discussion}
\label{results}

\begin{figure}
\begin{center}
\parbox{8cm}{
\epsfxsize=8cm \epsfbox{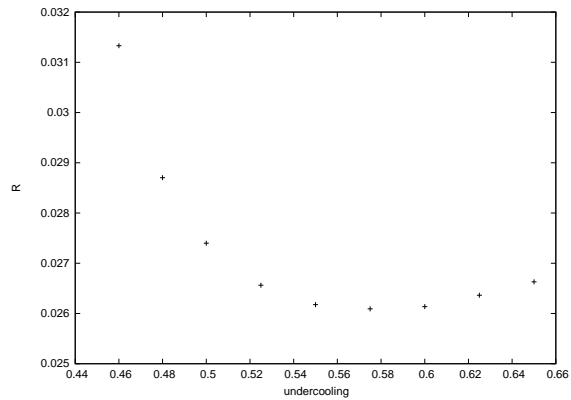}} \caption[]{\label{fig2} Tip radius of the
dendrite {\em vs.} undercooling.}
\end{center}
\end{figure}

A first series of simulations have been performed exploring the effect of
undercooling on the tip radius. This has been measured by computing
$R=\phi_{x}/\phi_{yy}$ at the tip of the dendrites \cite{wheeler}.
The aim has been to identify different regimes of growth in a large range of
$\Delta$.
In our case, anisotropy is only considered in surface tension. Thus, when the
undercooling is changed no change in growth direction
is
expected although the behavior of the tip radius and velocity may vary. In
particular in the surface tension controlled regime the tip radius should decrease
(together with increase of tip velocity) by increasing undercooling. On the
contrary in the kinetic regime, with isotropic kinetic term, one expects larger
tip radius at higher velocities. In Fig.~\ref{fig2} it is shown the behavior of
the tip radius as a function of the undercooling.
It can be clearly observed a change of the behavior around the value 0.575. Thus,
by choosing appropriate values of $\Delta$ we can select both regimes of growth.
The existence of these two different regimes can be also confirmed by looking at
the behavior of the tip velocity as a function of the undercooling or the P\`eclet
number.

We have looked at the shape of the studied dendrites by computing the
coordinates $(X,Z)$, as defined above following Ref. \cite{li98}. Thus, only data
of active branches are taken into account, {\it i.e.} branches longer than any
other closer to the tip.
\begin{figure}[h]
\begin{center}
\parbox{8cm}{
\epsfxsize=8cm \epsfbox{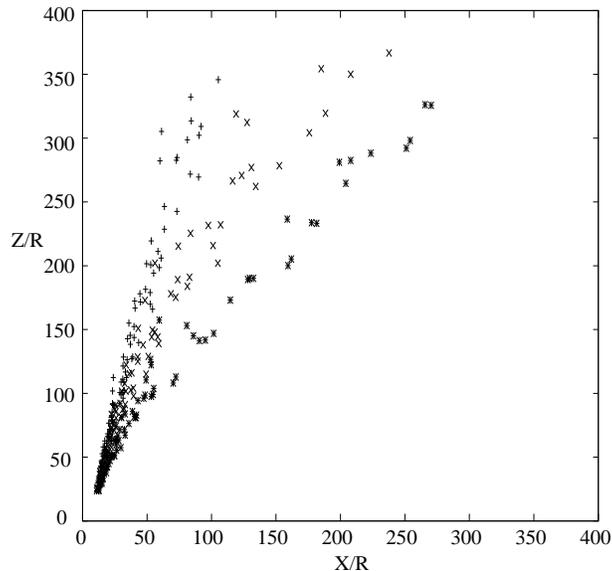}} \caption[]{\label{fig3} Plot of $Z/R$ {\em
vs.} $X/R$ for the sidebranches which are larger than any others closer to the
tip. Symbols $+$, $\times$ and $\ast$ correspond to $\Delta=0.48,0.55$ and 0.625,
respectively.} \end{center}
\end{figure}

Fig.~\ref{fig3} shows the plot of $Z/R$ {\it vs.} $X/R$ for three different values
of the undercooling ($\Delta=0.48,0.55,0.625$). Representation for each $\Delta$
contains data from eight different times, which explains the slight dispersion of
points. It can be distinguished in Fig.~\ref{fig3} the existence of two regimes
for each undercooling, which is more evident as the undercooling is increased. For
small values of $Z/R$, similar behaviors are found for all the undercoolings.
However, from a certain value of $Z/R$, $X/R$ depends very much on $\Delta$. The
transition region between both regimes is not clear enough to permit the precise
location of a crossover point in this figure.

\begin{figure}[h]
\begin{center}
\parbox{8cm}{
\epsfxsize=8cm \epsfbox{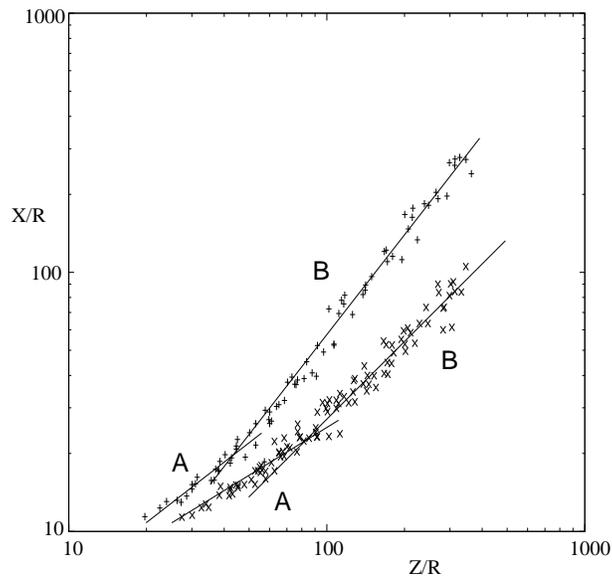}} \caption[]{\label{fig4} Log-log plot of
$X/R$ {\em vs.} $Z/R$ for the sidebranches which are larger than any others closer
to the tip. Symbols $\times$ and $+$ correspond to $\Delta=0.48$ and $0.65$,
respectively. In each case, regions A and B are indicated.}
\end{center}
\end{figure}

In order to better characterize the two observed regimes, it is shown in
Fig.~\ref{fig4} the log-log plot of $X/R$ {\it vs.} $Z/R$ for the smallest and the
largest undercoolings presented in Fig.~\ref{fig3}.
It can be observed from Fig.~\ref{fig4} a clear change in the behavior of $X/R$
in the regions of $Z/R$ around 80 and 40 for the case of small and large
undercoolings, respectively. 
This suggests that these regions separate two zones A and B (see
Fig.~\ref{fig4}) where sidebranching is in different regimes. For $\Delta \le
0.48$ it is difficult to distinguish these two regimes because their slopes are
very similar and it is not possible to determine a transition region. When the
undercooling is increased it is found that the transition region is closer to the
tip, which is consistent with the fact that at larger undercoolings 
a more developed sidebranching is obtained. 

The behavior of $X/R$ in region A is not exactly the same for all the
considered undercoolings and is given by a straight line in the log-log plot. By
comparing data from different undercoolings, it can be observed that the set of
points in region A lies at larger values of $X/R$ in the case of larger $\Delta$.
This is consistent with observations reported in Ref.~\cite{courgc1,courgc2},
where the exponents $a$ calculated for each single branch in $x \sim t^a$, $x$
being branch length and $t$ being time, were systematically smaller in branches
grown in lower undercooling conditions. According to this, at any value of the
undercooling and at small $Z/R$, points in $X/R(Z/R)$ representation are less
dispersed than at large $Z/R$ because in the region closer to the tip it is still
too soon to see the effects of the difference in the exponent $a$ of each branch.
Thus, the dispersion of points must increase with $Z/R$, as can also be observed
in Fig.~\ref{fig3}.

\begin{figure}[h]
\begin{center}
\parbox{8cm}{
\epsfxsize=8cm \epsfbox{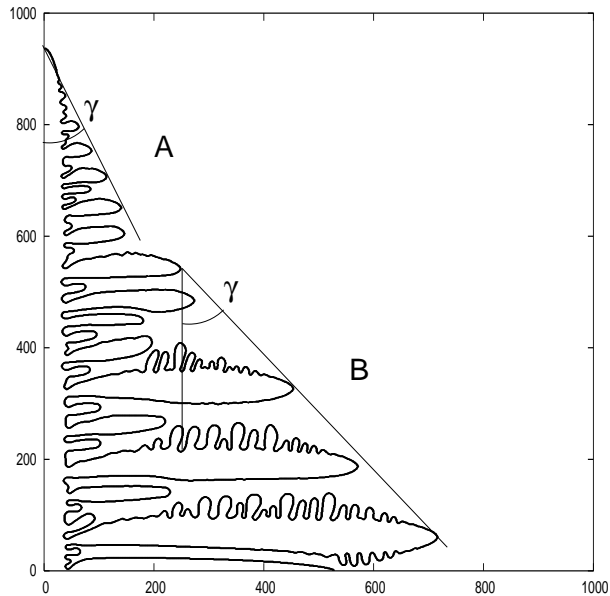}} \caption[]{\label{fig5} Dendrite grown at
$\Delta=0.6$ where regions A and B are indicated. Angle $\gamma$ is always larger
in regions B, where its value tends to 45 degrees as the undercooling is
increased.} \end{center}
\end{figure}

The behavior of $X/R$ in region B depends on $\Delta$. When fitting the points of
region B to $X/R \sim (Z/R)^{\alpha}$ we found that values of $\alpha$ tended to
1 for increasing $\Delta$ (in fact, from $\Delta \ge 0.525$, variations in
$\alpha$ are very small).
In Fig.~\ref{fig5} it is shown a dendrite grown at $\Delta=0.6$ where regions A
and B can be clearly distinguished. It can be observed that the angle $\gamma$
formed by the line joining the tips of active sidebranches and the axis of the
main arm is smaller in the region A and, typically for this range of
undercoolings, its value is very close to 45 degrees in the region B.
In other words, in region B sidebranches grow at the same velocity that the main
tip, {\it i.e.} grow as free dendrites. 

The observation of these two regimes reveals a significant difference with
experiments presented in Ref. \cite{li98} (see Fig. 7 there), where only one
regime was observed, and the angle formed by the axis of the main arm and the line
joining the tips of the branches were always considerably smaller than 45 degrees.
This has to be related to the small values of undercooling used in these
experiments. The diffusion lengths associated with such a slow growth are very
large, and even in the furthest from the tip region considered in the
experiments the process of competition between branches was not finished yet.  On
the contrary, diffusion length in simulations is short due to the large
undercoolings used. Branches can grow as free dendrites  as long as distances
between active sidebranches (which increase with the distance to the tip due to
the competition process) are larger than the interaction scale given by the
diffusion length (which is reduced for larger growth velocities). This results to
be the condition for the zone B to be observed.

We have also measured the integral parameters (contour length $U$ and area $F$)
of our two-dimensional dendrites. As shown in Fig.~\ref{fig1}, $U$ is the length
of the contour of the dendrite measured from the tip to a distance $Z$ along the
axis, while $F$ is one half of the area of the dendrite. Both magnitudes have been
measured for different values of the dimensionless undercooling.
Considering the origin of coordinates at the tip, the contour length and the area
have been calculated from the coordinates of the dendrite contour by %
\begin{equation}
U=\sum_{i=1}^{n}[(Z_{i+1}-Z_{i})^{2}+(X_{i+1}-X_{i})^{2}]^{1/2} \end{equation} %
and
\begin{equation}
F=\sum_{i=1}^{n} \frac{(X_{i+1}+X_{i})}{2}(Z_{i+1}-Z_{i}),
\end{equation}
where $n$ corresponds to each distance to the tip for which we calculated $U$ and
$F$. The fact that the shape of sidebranches is rather irregular and that their
growth is not always perpendicular to the $y$-axis, makes it difficult to define
U and F in a unique way everywhere as a function of $Z$. In order to better define
both functions and following Ref. \cite{li98}, we have only considered U and F for
the values of $Z$ corresponding to valleys between two neighboring sidebranches.

\begin{figure}[h]
\begin{center}
\parbox{8cm}{
\epsfxsize=8cm \epsfbox{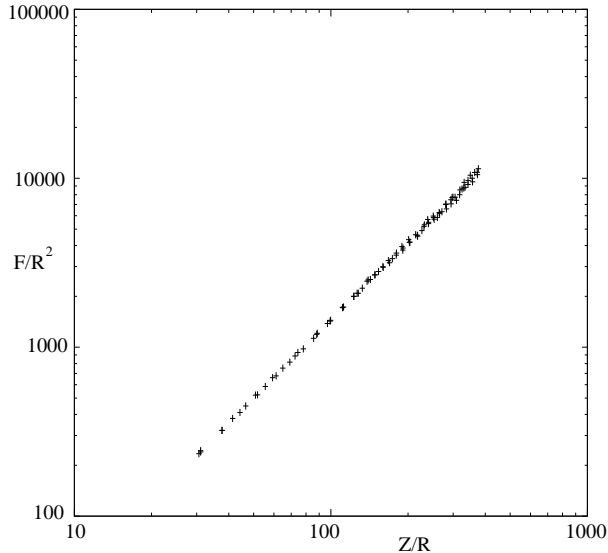}} \caption[]{\label{fig6}  Log-log plot of
$F/R^2$ {\it vs.} $Z/R$, F being the surface area, for $\Delta=0.525$.}
\end{center}
\end{figure}

It is shown in Fig.~\ref{fig6} the log-log plot of the normalized value of the
surface area as a function of the normalized value of the distance to the tip of
the dendrite at $\Delta=0.525$ with data taken at three different times. The same
representation for the rest of the considered undercoolings shows the same
behavior and only points far from the tip at larger $\Delta$ are slightly
dispersed.

$F/R^{2}$ {\em vs.} $(Z/R)$ follows a power-law ($a \sim b^{c}$) where $c$ is
always around 1.5, although a slight tendency to increase with $\Delta$ is also
observed. The value of $c$ found in the simulations completely coincides with that
found in Ref. \cite{couder} for the growth of ammonium bromide crystals in two
dimensions. However, in Ref. \cite{li98,li99} the representation of
$F/R^{2}(Z/R)$ showed two regimes of power-law behavior with different
exponents. In principle, we should not expect to find the same exponent in our
simulations taking into consideration that these experiments were
three-dimensional and it was measured the projection area of the dendrite. In this
case, the existence of different regimes near and far from the tip of the dendrite
was attributed to the different effect of coarsening. This effect is also present
in two dimensions, but the fact that only one regime is observed in the plot of
$F/R^{2}$ makes us conclude that the manifestation of the
coarsening effect is less dramatic in 2D that in 3D. As regards to the prefactor
of the power-law fitting, it shows a similar
behavior to that of the tip radius, that is, it decreases when the undercooling
is increased up to $\Delta < 0.55$ and it increases for larger $\Delta$.

\begin{figure}[h]
\begin{center}
\parbox{8cm}{
\epsfxsize=8cm \epsfbox{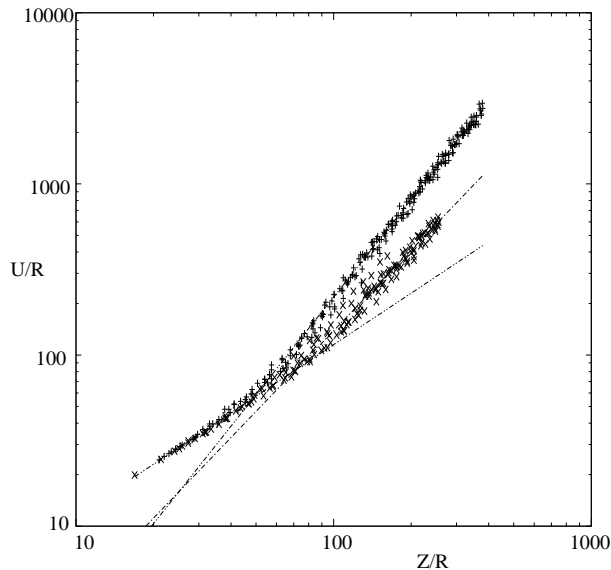}} \caption[]{\label{fig7}  Log-log plot of
$U/R$ {\it vs.} $Z/R$, U being the contour length. Symbols x and + correspond to
$\Delta=0.44$ and $0.6$, respectively. Lines indicate the fitting of points in
each region.} \end{center}
\end{figure}

The behavior of the normalized contour length as a function of the distance to the
tip for $\Delta=0.44$ and $0.6$ is shown in Fig.~\ref{fig7}. As it happened in the
plot of the shape of the dendrite, the behavior changes after a
transition region, being the variation of $U/R$ larger in the regions further down
from the tip.

In the region closer to the tip, it is found $U/R \sim (Z/R)^{1}$, which
coincides with the behavior found in the linear regime of experiments in Ref.
\cite{li99}. As one should expect, simulation results show that the linear region
is larger for smaller undercoolings. This was not observed in Ref. \cite{li99},
probably due to the employed range of undercoolings.

After the linear region, there is a transition region which is followed by the
nonlinear region, as it was observed in the experiments (see Fig. 5 in Ref.
\cite{li99}). The transition region in the $U/R(Z/R)$ plot is at smaller values
of $Z/R$ than the transition region in the $X/R(Z/R)$ plot (Fig.~\ref{fig4}) for
all the considered undercoolings. In fact, the change in the behavior of the
contour length and that of the shape of the dendrite (the envelop of it) provide
different information of the sidebranching activity.
In the case of $U$, the linear and nonlinear regimes are associated to low and
high developed perturbations of the interface respectively. In the nonlinear
region, both active and non-active branches contribute to the calculation of $U$.
The enhanced growing of the active branches observed in the nonlinear region is
accompanied by coarsening, process in which the shrinking of the shorter
branches reduces the total increasing of the contour length.
As a result any nonlinearity has effect on the behavior of $U$.
On the contrary the shape $X/R(Z/R)$ is calculated through the active branches
only. $X/R(Z/R)$ is then associated to the effect on the winning branches of
competition, and to these larger branches reaching colder regions and hence
growing faster. As a result, dispersions of $X/R$ values are rather large, and the
change of behavior more difficult to locate with a tendency to occur inside the
nonlinear region.
The behavior of $U/R$ in the nonlinear region can be fitted by a power-law (see
Fig.~\ref{fig7}), although the values of the prefactor and the exponent depend on
the undercooling. As regards to the prefactor, its variation with $\Delta$ is very
similar to that of the tip radius, that is, it decreases up to $\Delta \sim 0.55$
(surface tension dendrites) and it increases at larger undercoolings   (kinetic
dendrites). The largest value of the prefactor is $0.101$ for
$\Delta=0.44$, which is very far from the value obtained in Ref. \cite{li99}. The
divergence is probably related to the different ranges of undercoolings used in
simulations and experiments, but the difference in dimensions could also play a
role.

As regards to the exponent in the fitting of $U/R$, it increases with the
undercooling from $1.57$ to $1.89$. The value for small $\Delta$ is very similar
to the unique value ($1.50$) obtained in experiments \cite{li99}. Again, the fact
of having found many exponents in the simulations and only one in the experiments
could be associated to the different ranges of $\Delta$ used. It implies that
diffusion length considerably vary between simulations and
experiments. The influence of the diffusion length on the competition process
between branches that takes place in the nonlinear region determines the
evolution of branches and consequently the behavior of the contour length.

By combining results of the integral parameters in the linear regime, it is found
$F/(UR) \sim (Z/R)^{0.5}$, which coincides with the experiments
\cite{li99}. The same exponent in the nonlinear regime varies from $-0.07$ to
$-0.39$, differing very much from the experiments. Thus, the similarities between
our results and the experimental ones in Ref. \cite{li99} remain mainly in the
linear region and in the opposition to previous studies
\cite{hurlimann,li98}, where $F/UR$ was found to be constant in the nonlinear
regime.

\section{Concluding remarks}\label{conc}

We have presented a numerical study of the shape and sidebranching in regions at
different distances from the tip of a solidifying dendrite by means of a
phase-field model with a non-conserved noise term. We have characterized the
dendrite by using the integral parameters and we have focussed in dendrites grown
in both surface tension and kinetic regimes.

The behavior of the shape of the dendrite has been found to depend on the
undercooling in the considered range. The different diffusion lengths make the
competition process between sidebranches to differ and thus the final shape of the
dendrite is affected. The region where the competition process is taken place (A)
and that where it is finished and sidebranches evolve like free dendrites (B) have
been clearly distinguished in our simulation results. The behavior observed in B
is in agreement with theoretical predictions
\cite{brener95}. On the other hand, the main divergence with the available
experiments \cite{li98} is precisely the existence of two behaviors in the
nonlinear region. This discrepancy may be explained by the different range of
undercoolings considered in experiments and simulations. In our simulations
undercooling (and hence tip velocity) is larger, so diffusion length is smaller
and the transition between both zones is expected to occur closer to the tip,
becoming observable. Note that additional increasings of undercooling would reduce
further the size of zone A, which then could not be considered as a separate
scaling region.
The area of the dendrite presented a unique behavior for all the considered
undercoolings. As one would expect, it coincides with that of two-dimensional
dendrites \cite{couder}, although there is a slight discrepancy with
three-dimensional dendrites, especially in regions far from the tip. The
behavior $F/R^2 \sim (Z/R)^{1.5}$ found in our simulations is the same as if the
dendrites had a smooth parabolic shape. Thus, we can conclude that the area of
two-dimensional dendrites is
basically independent of the appearance and competition of sidebranches, and that
situation does not depend of the diffusion length of the system.

The behavior of the contour length presents two differentiated regimes in the
linear and nonlinear regions. The exponents found in the linear region are in
agreement with experiments \cite{li99}, while the discrepancies appeared in the
nonlinear region could come from the range of undercoolings or the
dimensionality.

We have found that the behavior of the contour length changes in regions closer
to the tip than the behavior of the shape does. The picture is the
following: As one moves down from the tip, one first finds the linear region where
branches are born and eventually start to compete with eachother. Going further,
the nonlinear region appears after a transition region. The competition process
is not only still taking place there but it is probably in the highest point of
activity. Not so far, its effects will be easily seen by the
observation of some already stopped sidebranches. During all this way we have
moved from the linear to the nonlinear region, but we are still in the region we
called A. Further down, the competition process between branches is finished and
the surviving ones have no opposition in their neighborhood to keep growing as
free dendrites. We are then in region B.

Finally, we have considered both surface tension and kinetic dominated
dendrites. Although the different behaviors of the studied parameters are observed
when the undercooling is changed, this cannot be associated to the type of
dendrite. What we can only assure is that the linear and A regions in surface
tension dendrites will always be larger than in the kinetic ones, but only because
of the larger diffusion length and not because of the main mechanism which
determines them.

These results offer some new insight into the understanding of a fully developed
dendrite, and in particular are of relevant importance to distinguish between low
and high undercooling dendrites and two-dimensional and three-dimensional
dendrites. Finally we should remark that it would be of the most great interest
to have more experimental results available in the high undercooling regime, in
particular characterizing the nonlinear regions of the dendrite.

\section*{Acknowledgments}

Authors thank discussions with Y. Couder. This research is supported by the
Direcci\'on General de Investigaci\'on Cient\'{\i}fica y T\'ecnica (Spain)
(Project BFM2003-07850-C03-02) and 
Comissionat per a Universitats i Recerca (Spain)
(Project 2001SGR00221).

\end{document}